\documentclass[
 aip,
 amsmath,amssymb,
 reprint,
]{revtex4-2}

\usepackage{graphicx}
\usepackage{dcolumn}
\usepackage{bm}
\usepackage[utf8]{inputenc}
\usepackage[T1]{fontenc}
\usepackage{mathptmx}
\usepackage{etoolbox}
\usepackage[per-mode=symbol]{siunitx}
\newcommand{\tblnum}[1]{\num[round-mode=places,round-precision=3]{#1}}

\makeatletter
\def\@email#1#2{%
 \endgroup
 \patchcmd{\titleblock@produce}
  {\frontmatter@RRAPformat}
  {\frontmatter@RRAPformat{\produce@RRAP{*#1\href{mailto:#2}{#2}}}\frontmatter@RRAPformat}
  {}{}
}%
\makeatother
\begin{document}

\title[Dynamic model of tissue electroporation on the basis of biological dispersion and Joule heating]{Dynamic model of tissue electroporation on the basis of biological dispersion and Joule heating}

\author{R. Guedert}
\email{raulguedert@gmail.com}
\affiliation{Institute of Biomedical Engineering, Federal University of Santa Catarina, Florian\'opolis, SC, Brazil}

\author{D. L. L. S. Andrade}
\affiliation{Institute of Biomedical Engineering, Federal University of Santa Catarina, Florian\'opolis, SC, Brazil}

\author{J. R. Silva}
\affiliation{Institute of Biomedical Engineering, Federal University of Santa Catarina, Florian\'opolis, SC, Brazil}

\author{G. B. Pintarelli}
\affiliation{Department of Control, Automation and Computer Engineering, Federal University of Santa Catarina, Blumenau, SC, Brazil}

\author{D. O. H. Suzuki}
\affiliation{Institute of Biomedical Engineering, Federal University of Santa Catarina, Florian\'opolis, SC, Brazil}

\date{\today}

\begin{abstract}
Electroporation is a complex, iterative, and nonlinear phenomenon that is often studied by numerical simulations. In recent years, tissue electroporation simulations have been performed using static models. However, the results of a static model simulation are restricted to a fixed protocol signature of the pulsed electric field. This paper describes a novel dynamic model of tissue electroporation that also includes tissue dispersion and temperature to allow time-domain simulations. We implemented the biological dispersion of potato tubers and thermal analysis in a commercial finite element method software. A cell electroporation model was adapted to account for the increase in tissue conductivity. The model yielded twelve parameters, divided into three dynamic states of electroporation. Thermal analysis describes the dependence of tissue conductivity on temperature. The model parameters were evaluated using experiments with vegetal tissue (\textit{Solanum tuberosum}) under electrochemotherapy protocols. The proposed model can accurately predict the conductivity of tissue under electroporation from \qty{10}{\kilo\volt\per\meter} to \qty{100}{\kilo\volt\per\meter}. A negligible thermal effect was observed at \qty{100}{\kilo\volt\per\meter}, with a \qty{0.89}{\celsius} increase. We believe that the proposed model is suitable for describing the electroporation current on a tissue scale and also for providing a hint on the effects on the cell membrane.
\end{abstract}

\maketitle

\section{Introduction}

Electroporation occurs when the transmembrane potential (TMP) of a biological cell exceeds a supraphysiological threshold by stimulation of short but intense pulsed electric fields (PEF). Excessive TMP causes local disturbances in the membrane structure. Current electroporation theory suggests that prepores are formed~\cite{Kotnik_2012}. The prepores then expand and stabilise as hydrophilic pores. Pore formation increases cell membrane permeability, allowing non-permeable substances to cross the cell barrier \cite{chen2020electroporation}. Extracellular content can access the cytosol, or intracellular content can leak and even trigger apoptosis by losing homeostasis \cite{napotnik2021cell, jakstys2020correlation}. There are medical and industrial applications that use electroporation to improve or replace traditional processes~\cite{Haberl_2013, Rakoczy_2022}.

The occurrence of electroporation depends on the distribution of the electric field, which in turn depends on the nonlinear electrical properties of the tissue. Opening the pores in the membrane changes the structure of the material and its electrical properties \cite{delemotte2012molecular}. The system is interdependent and leads to a complex model. For this reason, electroporation is often studied in detail by computer simulations.

Electrochemotherapy is a well-known application of electroporation to catalyse the membrane transport of chemotherapeutic drugs~\cite{Campana_2019}. The technique relies on standard protocols to ensure that the entire tumour is exposed above a minimum electric field threshold for electroporation to occur. The European Standard Operating Procedures for Electrochemotherapy (ESOPE) recommend a burst of eight rectangular pulses (monopolar, bipolar, or alternating) \qty{100}{\micro\second} long with a repetition rate between \qty{1}{\hertz} and \qty{5}{\kilo\hertz} \cite{Mir_2006, Gehl_2018}. Electrochemotherapy studies usually do not focus on the dynamics of pore formation, but rather on the final electric field distribution and outcomes. Static models are often used to reduce computational costs~\cite{Martins_Taques_2020, Andrade_2022}. However, a static electroporation model is developed specifically for a PEF signature and cannot be directly used to study different signatures. If the PEF signature is changed, the static model needs to be adjusted. The reason is that a PEF has a specific energy spectrum density that affects the dynamics of electroporation, Joule heating, and the dispersive aspect of biological media~\cite{Guedert_2023}. To overcome this limitation, dynamic models should be used.

A dielectric dispersive medium is characterised by the dependence of its electrical properties on frequency. Biological tissue exhibits a strong dispersion from DC to hundreds of \si{\giga\hertz}. There are four main dispersion bands in biological tissue: $\alpha$ (from DC to about \qty {10}{\kilo\hertz}), $\beta$ (between \qty{100}{\kilo\hertz} and \qty{10}{\mega\hertz}), $\gamma$ (at \qty{20}{\giga\hertz}), and $\delta$ (between $\beta$ and $\gamma$). Tissue electroporation uses square-wave PEF with a broad spectral distribution. For this reason, including biological dispersion is essential to accurately simulate the biological medium. Then a dynamic model of tissue electroporation should be developed based on biological dispersion~\cite{Guedert_2023, Guedert2023-2}.

Proposals for tissue dynamic models have already been published~\cite{Neal_2012, Suarez_2014, Zhao_2020, Langus_2016, Voyer_2018, Ramos_2018, Weinert_2019, Weinert_2021}. Some are for the analysis of irreversible electroporation, considering the dynamics of the temperature rise and its effects on the tissue conductivity. However, they do not consider time-dependent changes in electrical properties due to electroporation~\cite{Neal_2012, Suarez_2014, Zhao_2020}. On the other hand, three models consider time-dependent changes in electrical properties due to electroporation~\cite{Langus_2016,Voyer_2018,Ramos_2018, Weinert_2019, Weinert_2021}. Yet, there are some modelling simplifications. One did not consider the biological dispersion \cite{Langus_2016}, while the other two consider the tissue dispersion to be $\beta$-based~\cite{Voyer_2018, Ramos_2018, Weinert_2019, Weinert_2021}. Nevertheless, it is known that for most electroporation PEF protocols, more than 95\% of the spectral energy is below \qty{100}{\kilo\hertz}~\cite{Guedert_2023}, which is in the $\alpha$-dispersion band. Although the three models~\cite{Voyer_2018, Ramos_2018, Weinert_2019, Weinert_2021} can accurately describe the shape of the electric current during tissue electroporation, some of their parameters are adjusted according to the experimental setup, such as voltage and shape of the electrode. These adjustments hamper the use of models if the experimental setup is changed, which is common practice in PEF research. Thus, the input parameters should be redefined. Furthermore, the three models were developed using custom numerical solution software and not an immediate implementation.

In this paper, we present a novel dynamic model of tissue dielectric properties during PEF that accounts for the three individual physical effects: electroporation, tissue dispersion, and temperature. We describe tissue dispersion using a multipole Debye function implemented in the time domain by the method of auxiliary differential equations. The electroporation effect was described using an electric-field-to-TMP relation. The time-dependent increase in tissue conductivity was evaluated by extrapolating a kinect model of cell electroporation. In addition, the conductivity dependence on temperature was included. We used \textit{in vitro} potato tuber (\textit{Solanum tuberosum}) to collect data and implement the model using commercial finite element method (FEM) software.

\section{Methods}

\subsection{Model Development}
We solve the models using numerical simulations. We used the commercial software COMSOL Multiphysics (COMSOL Inc., Stockholm, Sweden). COMSOL is both a FEM and a computational fluid dynamic (CFD) solver with a bundle of built-in physical equations. We used COMSOL's electric current library, which solves electrophysics for low-frequency signals using the FEM. In low-frequency electrophysics, COMSOL use the principle of charge conservation and solve the equation
of continuity shown in Eq.~\ref{eqn:continuity-equation} and the electric current density given by Eq. \ref{eqn:current-density}.

\begin{equation}\label{eqn:continuity-equation}
    \vec{\nabla} \cdot \vec{J}(t) = Q_j
\end{equation}

\begin{equation}\label{eqn:current-density}
    \vec{J}(t) = \sigma \vec{E}(t) + \frac{\partial \vec{D}(t)}{\partial t} + \vec{J}_e(t)
\end{equation}

\noindent where $Q_j$ is the total charge density (\si{\coulomb\per\cubic\meter}) and $\vec{J}(t)$ is the electric current density (\si{\ampere\per\meter\squared}) whose components are given by the material conductivity $\sigma$ (\si{\siemens\per\meter}), the electric field $\vec{E}$ (\si{\volt\per\meter}) and the displacement field $\vec{D}$ (\si{\coulomb\per\meter\squared}). $\vec{J}_e$ is an arbitrary external electric current density assumed by COMSOL to allow inclusion of external effects. The Maxwell equations define the displacement field as shown in Eq.~\ref{eqn:electric-displacement-vector}, where $\epsilon_0$ is the vacuum permittivity and $\varepsilon_r$ is the relative permittivity of the material.

\begin{equation} \label{eqn:electric-displacement-vector}
    \vec{D}(t) = \epsilon_0 \varepsilon_r \vec{E}(t)
\end{equation}

\subsubsection{Biological Dispersion}

A biological medium is a dielectric that has regions susceptible to polarisation and consequently to charge relaxation times. In the frequency domain, the relaxation effect is called dispersion and can be represented as a complex relative permittivity function. There are several models for describing the dispersion behaviour in biological materials. The Debye dispersion can be implemented in the time domain using a set of auxiliary differential equations~\cite{Guedert2023-2}. Eq.~\ref{eqn:debye-dispersion} represents the multipole Debye model in the frequency domain.

\begin{equation}\label{eqn:debye-dispersion}
     {\varepsilon}_{r}^{*} (\omega) = \varepsilon_\infty + \frac{\sigma_s}{j \omega \epsilon_0} + \sum_{k=1}^{N} \frac{\Delta \varepsilon_k}{1 + \left( j \omega \tau_k \right)}
\end{equation}

\noindent where $\sigma_s$ is the static conductivity, $\omega$ is the angular frequency (\si{\radian\per\second}), $\varepsilon_\infty$ is the permittivity of the material at high frequency, $\Delta \varepsilon_k$ is the permittivity variation of the pole, and $\tau_k$ is the relaxation time of the pole (\si{\second}). $k$ and $N$ are the current pole number and the total number of poles, respectively. $j$ is the imaginary unit.

The implementation of the multipole Debye model in the time domain followed our previously proposed method~\cite{Guedert2023-2}, in which the dispersive effect is contained in an external electric current density for each Debye pole ($\vec{J}_{e_k}$), as shown in Eq.~\ref{eqn:current-density-debye}. Each current density is solved using an auxiliary electric field ($\vec{e_k}$) as shown in Eq.~\ref{eqn:external-current-debye}, which is a delayed value of the input electric field ($\vec{E}$) with a time constant corresponding to the relaxation of the Debye pole as shown in Eq.~\ref{eqn:auxiliary-electric-field-debye}. This set of equations is implemented in COMSOL Multiphysics with the domain ordinary differential equation (DODE) physics.

\begin{equation}\label{eqn:current-density-debye}
    \vec{J}(t) = \sigma_s \vec{E}(t) + \epsilon_0\varepsilon_\infty \frac{\partial \vec{E}(t)}{\partial t} + \sum_{k=1}^{N} \vec{J}_{e_k}(t)
\end{equation}

\begin{equation}\label{eqn:external-current-debye}
    \vec{J}_{e_k}(t) =  \frac{\epsilon_0\Delta\varepsilon_k}{\tau_k} \left( \vec{E}(t) - \vec{e_k}(t) \right)
\end{equation}

\begin{equation}\label{eqn:auxiliary-electric-field-debye}
    \tau_k \frac{\partial \vec{e_k}(t)}{\partial t}  = \vec{E}(t) - \vec{e_k}(t)
\end{equation}

We have previously described the dielectric spectrum of \textit{Solanum tuberosum} tissue with different numbers of Debye poles~\cite{Guedert2023-2}. The dielectric dispersion from \qty{40}{\hertz} to \qty{10}{\mega\hertz} can be parameterised with a 4-pole Debye dispersion model. The parameters are shown in Table~\ref{tbl:debye-parameters}.

\begin{table}[htb]
\caption{\label{tbl:debye-parameters}Parameterisation of potato tissue dispersion with the 4-pole Debye dispersion model~\cite{Guedert2023-2}.}
\begin{ruledtabular}
    \begin{tabular}{lrlr}
    \textbf{Parameter}      & \textbf{Value}        &  \textbf{Parameter}      & \textbf{Value}          \\  \hline
    $\epsilon_{\infty}$     & \tblnum{1.746734E+02} &  $\tau_2$ (\si{\second}) & \tblnum{2.309457E-05}   \\ 
    $\sigma_s$              & \tblnum{2.158834E-02} &  $\Delta\varepsilon_{3}$    & \tblnum{1.836403E+04}   \\ 
    $\Delta\varepsilon_{1}$    & \tblnum{2.251149E+06} &  $\tau_3$                & \tblnum{1.005246E-06}   \\ 
    $\tau_1$ (\si{\second}) & \tblnum{3.783432E-03} &  $\Delta\varepsilon_{4}$    & \tblnum{1.052989E+04}   \\ 
    $\Delta\varepsilon_{2}$    & \tblnum{2.917574E+04} &  $\tau_4$ (\si{\second}) & \tblnum{1.658463E-07}   \\ 
    \end{tabular}
\end{ruledtabular}
\end{table}

\subsubsection{Electroporation}

Electroporation dynamics was described on the basis of a modified version of the Leguèbe \textit{et al.}~\cite{Leguebe_2014} cell model. We modelled membrane electroporation using three states: prepore ($P_0$), initial pore ($P_1$), and expanded pore ($P_2$). Each state contributes in a specific way to the increase in membrane conductivity. In the cell model, the pore state is related to the TMP. In a macro-tissue model, we cannot access TMP directly because we do not have the membrane geometric model. For this reason, we proposed to calculate the TMP based on the magnitude of the local electric field ($E$). The concentrations of the pore states grow and decay exponentially, $P_0$ and $P_1$ as a function of the electric field and $P_2$ as a function of $P_1$. The proposed pore formation system was described using a set of differential Eqs.~\ref{eqn:diff-p1} -- \ref{eqn:diff-L}. The notation in brackets indicates the concentration of the state.

\begin{equation}
\label{eqn:diff-p1}
\frac{\mathrm{d} [P_0]}{\mathrm{d}t}= \frac{\beta_0(E) - [P_0]}{\tau_{0}}
\end{equation}

\begin{equation}
\label{eqn:diff-p2}
\frac{\mathrm{d} [P_1]}{\mathrm{d}t}= \frac{\beta_1(E) - [P_1]}{\tau_{1}}
\end{equation}

\begin{equation}
\label{eqn:diff-L}
\frac{\mathrm{d} [P_{2}]}{\mathrm{d}t}=\frac{[P_1] - [P_{2}]}{\tau_{2}}
\end{equation}

$\tau_{0}$ is a constant time. $\tau_{1}$ and $\tau_{2}$ depend on whether the function is growing or declining, as represented in Eqs.~\ref{eqn:tau-1} and \ref{eqn:tau-2}. $\beta_0$ and $\beta_1$ describe the maximum concentration for the states $P_0$ and $P_1$ as a function of the magnitude of the electric field and are expressed in Eqs.~\ref{eqn:beta-0} and \ref{eqn:beta-1}, respectively. This means that both functions can vary between zero and one. Zero means that no prepore ($\beta_0$) or pore ($\beta_1$) is formed. One means that the tissue has reached saturation for each phenomenon. Note that saturation does not mean that no new prepores or pores are formed, but that their increasing number above a certain threshold no longer has significant influence on the conductivity of the tissue. The maximum number of pores is usually not considered in cell electroporation models. The asymptotic model \cite{neu1999asymptotic}, for example, does not define a limit value for the pore density.

\begin{equation}
\label{eqn:tau-1}
\tau_{1} = \left\{\begin{matrix}
\tau_{1G}(1 - 0.5[P_0]) & \text{if}\ \beta_1(E) - [P_1] \geq 0 \\
\tau_{1D} & \text{otherwise}
\end{matrix}\right.
\end{equation}

\begin{equation}
\label{eqn:tau-2}
\tau_{2} = \left\{\begin{matrix}
\tau_{2G} & \text{if}\ [P_1] - [P_2] \geq 0 \\
\tau_{2D} & \text{otherwise}
\end{matrix}\right.
\end{equation}

\begin{equation}
\label{eqn:beta-0}
\beta_0(E) = \frac{1}{1 + e^{-(|E| - E_0)/\Delta E_0}}
\end{equation}

\begin{equation}
\label{eqn:beta-1}
\beta_1(E) = \frac{1}{1 + e^{-(|E| - E_1)/\Delta E_1}}
\end{equation}

\noindent where $E_0$ and $E_1$ are the central values of each logistic function, and $\Delta E_0$ and $\Delta E_1$ shape the slope. $\tau_{1G}$ and $\tau_{2G}$ are the characteristic times of $[P_1]$ and $[P_2]$. $\tau_{1D}$ and $\tau_{2D}$ are the relaxation times of $[P_1]$ and $[P_2]$.

The increase in the number of pores ($P_1$) would probably decrease the number of prepores ($P_0$); the same is true for increasing the size of the pores ($P_2$) and their initial shape ($P_1$). We have not included this state interaction in our model definition because we are working with concentrations and not absolute numbers. It is not expected that all prepores lead to a pore and that all pores expand. According to~\cite{krassowska2007modeling}, about only 2.2\% of the pores will expand. We believe that its implementation would have little effect on the final result of the model while increasing the number of parameters.

Since we cannot directly deal with membrane conductivity in tissue simulation, the increase in conductivity was implemented in tissue dispersion. There are works that deal with the effects of electroporation on biological dispersion \cite{Yao_2017, Pintarelli_2022}. Impedance analysis before and after PEF stimulation showed an increase in tissue conductivity throughout the spectrum. Permittivity is also affected, but to a lesser extent. To increase conductivity across the spectrum, we increased the static conductivity of the biological dispersion ($\sigma_s$) according to Eq.~\ref{eqn:electroporation-conductivity}.

\begin{equation}\label{eqn:electroporation-conductivity}
    \sigma_P = \sigma_{s} + \sigma_{P_0}[P_0] + \sigma_{P_1}[P_1] + \sigma_{P_2}[P_2]
\end{equation}

\noindent where $\sigma_{P_0}$, $\sigma_{P_1}$, and $\sigma_{P_2}$ are the increasing coefficients of the states $P_0$, $P_1$, and $P_2$, respectively. $[P_0]$, $[P_1]$, $[P_2]$ are evaluated using Eqs.~\ref{eqn:diff-p1} -- \ref{eqn:diff-L}.

\subsubsection{Thermal Dependence}

The electrical conductivity of biological tissue increases with temperature~\cite{Brace_2008}. The electric current flowing through the tissue generates Joule heating. We simulated the temperature development in the sample during the pulse burst. The equation for heat diffusion with Joule heating term is presented in Eq.~\ref{eqn:heat-diffusion}.

\begin{equation}\label{eqn:heat-diffusion}
    \rho c_p \frac{\partial T}{\partial t} - \nabla \cdot \left( k \nabla T \right) = \vec{J} \cdot \vec{E}
\end{equation}

\noindent where $T$ is the temperature (\si{\kelvin}), $\rho$ is the density of the material (\si{\kilo\gram\per\cubic\meter}), $c_p$ is the heat capacity of the material at constant pressure (\si{\joule\per\kilo\gram\per\kelvin}), and $k$ is the thermal conductivity (\si{\watt\per\meter\per\kelvin}). $\vec{J}$ and $\vec{E}$ are the electric current density and electric field calculated with Eqs.~\ref{eqn:continuity-equation} and \ref{eqn:current-density}. The thermophysical properties of \textit{Solanum tuberosum}~\cite{Krishna_Kumar_2018} and the electrode used during the experiments (316L Stainless Steel~\cite{Trejos_Taborda_2022}) are given in Table~\ref{tbl:heat-parameters}.

\begin{table}[tb]
\caption{\label{tbl:heat-parameters}Thermophysical properties at \qty{20}{\celsius} for the materials used in the experiments.}
\begin{ruledtabular}
\begin{tabular}{lrrr}
\textbf{Material} & $\rho$ & $c_p$ & $k$ \\
\hline
\textit{Solanum tuberosum}~\cite{Krishna_Kumar_2018} & 1053 & 4410 & 0.56 \\ 
316L Stainless Steel~\cite{Trejos_Taborda_2022} & 8000 & 480 & 13.50 \\
\end{tabular}
\end{ruledtabular}
\end{table}

Temperature influences the increase in electroporation conductivity (Eq.~\ref{eqn:electroporation-conductivity}) because it affects all components of the tissue. The conductivity temperature coefficient was defined as $\chi$~=~\num{1.7E-2}~\si{\per\kelvin}~\cite{Neal_2012}. Thus, the conductivity is adapted as follows.

\begin{equation}\label{eqn:thermal-conductivity}
    \sigma_T = \sigma_{P} \left(1 + \chi (T - T_0) \right)
\end{equation}

\noindent where $T_0$ is the initial temperature.

After including thermal and electroporation effects, Eq.~\ref{eqn:current-density-debye} is adapted to Eq.~\ref{eqn:current-density-debye-effects} and used to describe the apparent conductivity in the simulator.

\begin{equation}\label{eqn:current-density-debye-effects}
    \vec{J}(t) = \sigma_T \vec{E}(t) + \epsilon_0\varepsilon_\infty \frac{\partial \vec{E}(t)}{\partial t} + \sum_{k=1}^{N} \vec{J}_{e_k}(t)
\end{equation}

\subsection{\textit{In vitro} experiment}

Potato tubers (\textit{Solanum tuberosum}) were brought from local growers. Growers were certified by the Brazilian Ministry of Agriculture, Livestock and Food Supply (MAPA) for organically grown products. Cylindrical fragments were cut using a \qty{18.50}{\milli\meter} diameter stainless steel cutter. The cylindrical fragments were then cut into \qty{5}{\milli\meter} tall samples. The samples were wrapped in paper towels to reduce denaturation and oxidation. Cutting was performed immediately before each experiment. The time between cutting and the end of the experiment was less than 30 minutes. The laboratory temperature was \qty{20}{\celsius} (\qty{293.15}{\kelvin}).

\begin{figure*}[htb]
\centering
\includegraphics{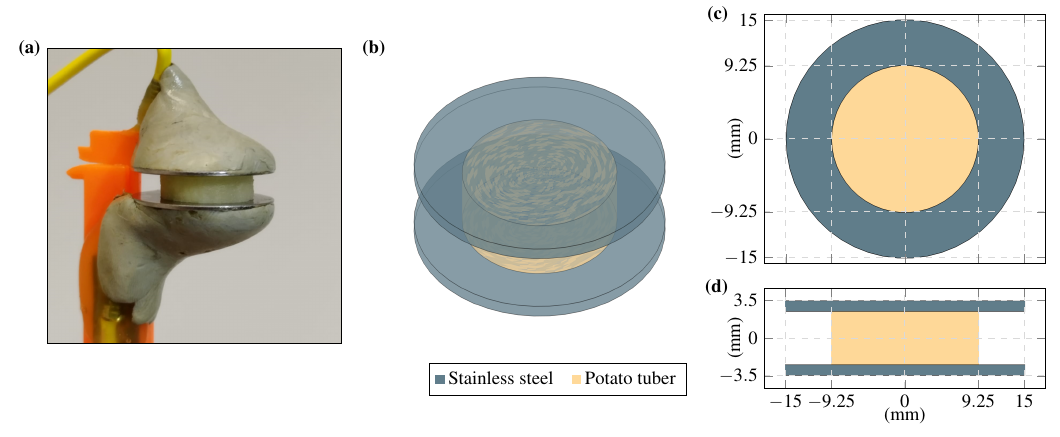}
\caption{The experimental sample and the rotated 2D axisymmetric geometry created in the simulator. \textbf{(a)} Experimental setup. \textbf{(b)} 3D representation. \textbf{(c)} Transversal view. \textbf{(d)} Frontal view. The geometry is centred at the origin.}
\label{fig:geometry}
\end{figure*}

The samples were placed between two 30 mm diameter circular 316L stainless steel plates as shown in Fig.~\ref{fig:geometry}a and carefully fixed with a spring clamp. We subjected the samples to a PEF protocol according to the ESOPE guidelines \cite{Gehl_2006}. The repetition rate was fixed at \qty{5}{\kilo\hertz}. Each sample was subjected to a PEF protocol and then replaced. The voltage was swept from \qty{50}{\volt} to \qty{250}{\volt} (\qty{50}{\volt} steps) and from \qty{300}{\volt} to \qty{500}{\volt} (\qty{100}{\volt} steps). Because the samples were \qty{5}{\milli\meter} high, the equivalent electric field ranged from \qty{10}{\kilo\volt\per\meter} to \qty{50}{\kilo\volt\per\meter} (\qty{10}{\kilo\volt\per\meter} steps) and from \qty{60}{\kilo\volt\per\meter} to \qty{100}{\kilo\volt\per\meter} (\qty{20}{\kilo\volt\per\meter} steps).

Data were collected using a Tektronix DPO2012B oscilloscope (Tektronix Inc, Oregon, USA) with a Tektronix TPP0100 voltage probe and a Tektronix A622 current probe. We post-processed the data using a Python script to determine the average, standard deviation, and confidence intervals for each protocol.

\subsection{\textit{In silico} experiment}

Iterative simulations were used to evaluate the parameters. We used a 2D axisymmetric geometry to replicate the experimental setup. The sample geometry was a rectangle \qty{9.25}{\milli\meter} long and \qty{5}{\milli\meter} high. Rectangles \qty{15.00}{\milli\meter} long and \qty{1}{\milli\meter} high were placed on the top and bottom of the sample geometry to form the plate electrodes. All geometries were rotated \qty{360}{\degree} to form the cylindrical shape. Fig.~\ref{fig:geometry}b shows the final geometry. We implemented the biological dispersion (Eqs.~\ref{eqn:current-density-debye} -- \ref{eqn:auxiliary-electric-field-debye}) with electroporation and thermal conductivities (Eqs.~\ref{eqn:thermal-conductivity} and \ref{eqn:current-density-debye-effects}) in the sample material. The conductivity and relative permittivity of the electrode were \qty{1.74}{\mega\siemens\per\meter} and 1, respectively.

The boundary conditions were defined as follows. A lateral boundary of the entire geometry was used for the axisymmetric rotation. For electrical analysis, the boundaries of one electrode were defined as terminal and those of the other as ground (Dirichlet boundary condition). For electrical and thermal analysis, the external boundaries of the geometry were defined as electrical and thermal insulating, respectively (Neumann boundary condition). The COMSOL's multiphysics module automatically provided the electrical information as input for thermal analysis, all domains were considered as Joule heating source. The initial temperature was \qty{20}{\celsius}.

The input voltage followed the experimental signal. The mesh was created with the COMSOL mesh creation tool using finer resolution. The final mesh resulted in 736 domain elements. We used the intermediate generalised-$\alpha$ method to adjust the time step to improve convergence. The intermediate generalised-$\alpha$ method allows the solver to strictly decrease the time step. The maximum time step was established at \qty{0.1}{\micro\second} in the transition regions and at \qty{1}{\micro\second} otherwise.

\section{Results}

\begin{figure}[tb]
\centering
\includegraphics{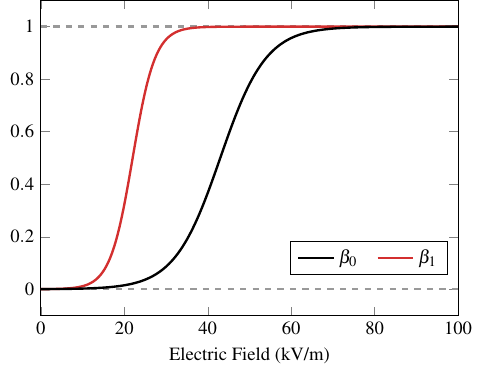}
\caption{Shape of sigmoidal functions $\beta_0$ and $\beta_1$ for potato tissue with parameters of Table~\ref{tbl:model-parameters}.}
\label{fig:beta}
\end{figure}

\begin{figure*}[htb]
\centering
\includegraphics{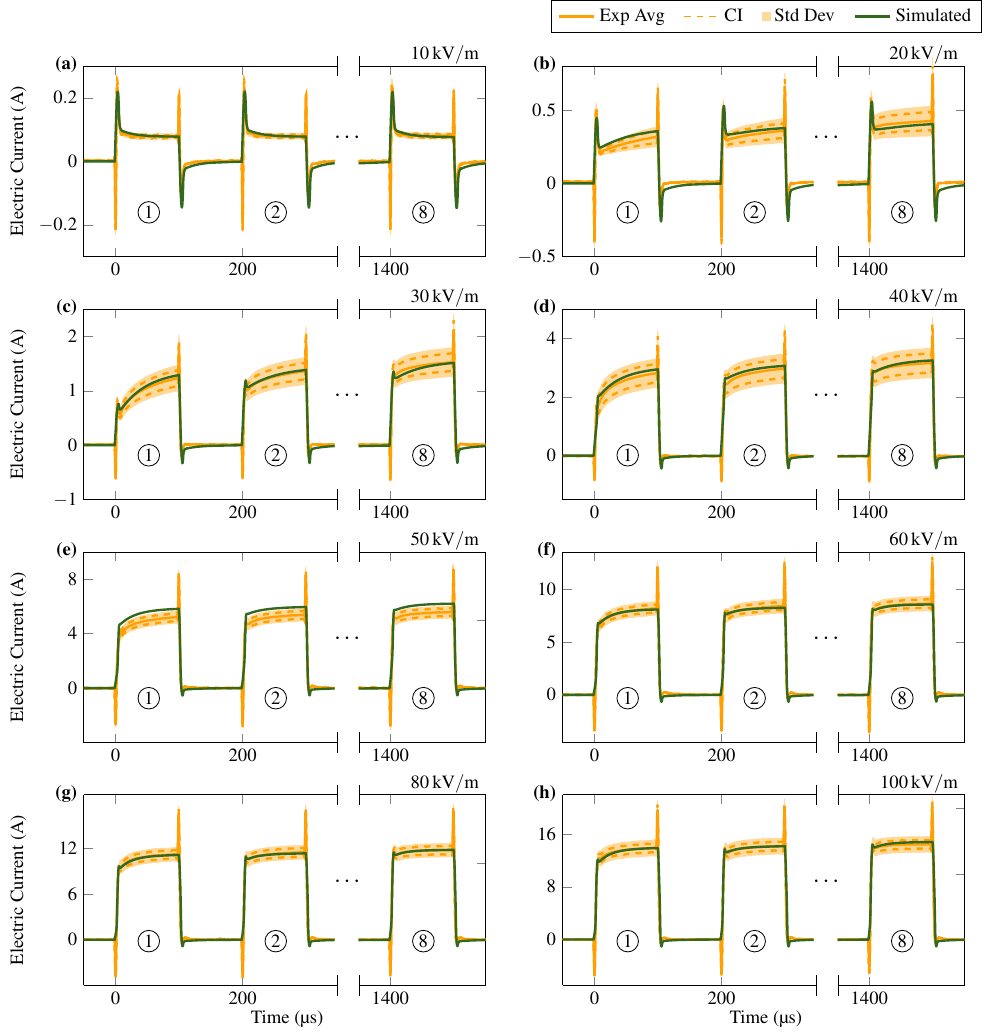}
\caption{Experimental and simulation results for the electroporation experiments following the ESOPE guidelines (eight pulses \qty{100}{\micro\second} long at \qty{5}{\kilo\hertz}). Only the first, second, and eighth pulses are shown; see Supplementary Information for the full set of pulses. Magnitude of the PEF protocol at \textbf{(a)} \qty{10}{\kilo\volt\per\meter}, \textbf{(b)} \qty{20}{\kilo\volt\per\meter}, \textbf{(c)} \qty{30}{\kilo\volt\per\meter}, \textbf{(d)} \qty{40}{\kilo\volt\per\meter}, \textbf{(e)} \qty{50}{\kilo\volt\per\meter}, \textbf{(f)} \qty{60}{\kilo\volt\per\meter}, \textbf{(g)} \qty{80}{\kilo\volt\per\meter}, and \textbf{(h)} \qty{100}{\kilo\volt\per\meter}. The overshoot of the \textit{in vitro} current in the opposite direction in the PEF transitions is a common parasitic effect in the experimental setup. The circled numbers indicate the pulse number. Exp Avg is the experimental average, CI is the confidence interval (95\%) and Std Dev is the standard deviation.}
\label{fig:electric-currents}
\end{figure*}

The parameters of the dynamic model of \textit{Solanum tuberosum} electroporation are shown in Table~\ref{tbl:model-parameters}. Fig.~\ref{fig:beta} presents the plot of the functions $\beta_0$ and $\beta_1$. Fig.~\ref{fig:electric-currents} shows the experimental and simulated electric currents for the used PEF (\num{10} to \qty{100}{\kilo\volt\per\meter}). The experimental results were summarised as average, confidence interval (95\%), and standard deviation. If PEF is higher than \qty{20}{\kilo\volt\per\meter}, the electric current has a nonlinear increase during PEF. Also, if PEF increases in magnitude, the maximum current values increase nonlinearly. The overshoot of the \textit{in vitro} current in the opposite direction in the PEF transitions is not due to dispersion, temperature rise, or electroporation, but is a common parasitic effect in the experimental setup.

\begin{table}[t]
\caption{\label{tbl:model-parameters}Electroporation parameters for potato tubers. Initial temperature was \qty{20}{\celsius}.}
\begin{ruledtabular}
    \begin{tabular}{lrlr}
    \textbf{Parameter} & \textbf{Value} & \textbf{Parameter} & \textbf{Value} \\
    \hline
    $E_0$ & 43 \si{\kilo\volt\per\meter} & $\tau_{1G}$ & 40 \si{\micro\second} \\
    $\Delta E_0$ & 5.5 \si{\kilo\volt\per\meter} & $\tau_{1D}$ & 150 \si{\micro\second} \\
    $E_1$ & 22 \si{\kilo\volt\per\meter} & $\sigma_{P_1}$ & 0.11 \si{\siemens\per\meter} \\
    $\Delta E_1$ & 2.7 \si{\kilo\volt\per\meter} &  $\tau_{2G}$ & 500 \si{\micro\second} \\
    $\tau_0$ & 0.5 \si{\micro\second} & $\tau_{2D}$ & 1 \si{\second} \\
    $\sigma_{P_0}$ & 0.375 \si{\siemens\per\meter} & $\sigma_{P_2}$ & 0.04 \si{\siemens\per\meter} \\
    \end{tabular}
\end{ruledtabular}
\end{table}

We evaluated the evolution of dynamic states and the thermal increase in the centre of the sample (coordinates (0;0) in Fig.~\ref{fig:geometry}c and d). The dynamic evolution of the concentrations of prepore ($P_0$), initial pore ($P_1$), and expanded pore ($P_2$) using \qty{20}{\kilo\volt\per\meter} and \qty{100}{\kilo\volt\per\meter} is shown in Fig.~\ref{fig:states} (see Supplementary Information for the complete set of input electric fields). $P_0$ are created and decimate during the first \qty{500}{\nano\second} after the pulse rise and fall times. $P_1$ are created at a rate faster than its decimation. A higher magnitude of the PEF leads to more pore formation. $P_2$ has the slower dynamics and accumulates over pulses.

\begin{figure*}[ht]
\centering
\includegraphics{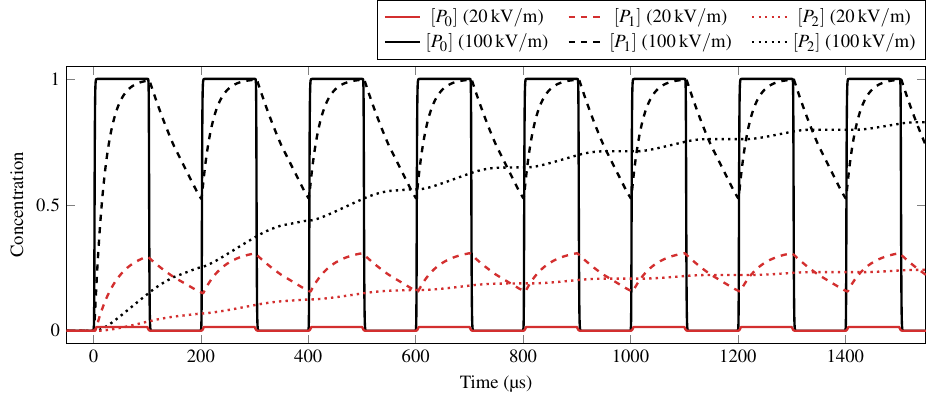}
\caption{Concentration of electroporation dynamic states $P_0$, $P_1$, and $P_2$ for \qty{20}{\kilo\volt\per\meter} and \qty{100}{\kilo\volt\per\meter} at the centre of the sample.}
\label{fig:states}
\end{figure*}

\begin{figure}[ht]
\centering
\includegraphics{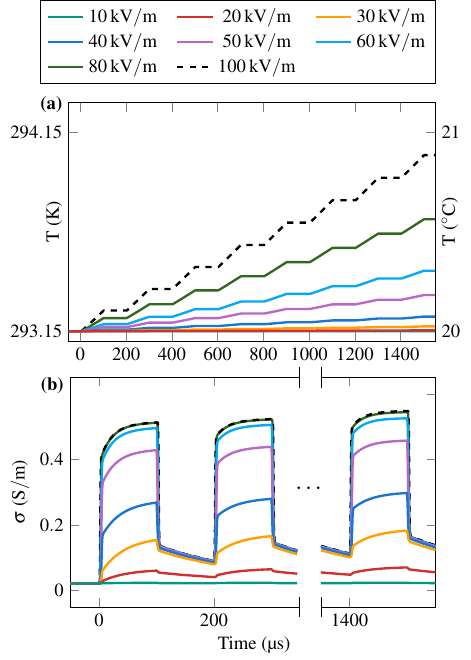}
\caption{\textbf{(a)} Temperature rise due to Joule heating at the centre of the sample. \textbf{(b)} Apparent conductivity with thermal and electroporation influences at the centre of the sample. For the apparent conductivity, only the first, second, and eighth pulses are shown; see Supplementary Information for the full set of pulses.}
\label{fig:thermal-and-conductivity}
\end{figure}

The increase in temperature is shown in Fig.~\ref{fig:thermal-and-conductivity}a. The total increase in temperature ($\Delta T = T - T_0$) after the eight pulses was \num{0}, \num{0.004}, \num{0.025}, \num{0.074}, \num{0.1822}, \num{0.3046}, \num{0.5647}, and \qty{0.8862}{\celsius} for \num{10}, \num{20}, \num{30}, \num{40}, \num{50}, \num{60}, \num{80}, and \qty{100}{\kilo\volt\per\meter}, respectively. The maximum temperature variation (\qty{0.8862}{\kelvin}) represents an increase in conductivity of 1.5\%.

Fig.~\ref{fig:thermal-and-conductivity}b presents the evolution of the apparent conductivity for all input PEF. Measurement was taken in the centre of the sample. This curve is calculated through Eq.~\ref{eqn:thermal-conductivity}, where the contribution of electroporation and thermal effects to tissue conductivity is included.

\section{Discussion}

Modelling complex biological systems is a challenging undertaking due to complex interactions, nonlinear dynamics, computational demand, and uncertainties inherent in these systems. Another difficulty is addressing parameters that correlate with the microphysical processes. Studies in single cells provide an opportunity to study the effects directly at the cell membrane \cite{Milestone2023}. On the other hand, tissue studies can only evaluate a macroscopic effect \cite{Andrade2023}. In electroporation, the increase in membrane conductivity (and thus the tissue conductivity) is one of the primary observable effects. Our tissue model subsumed the complex dynamics of electroporation into three main states $P_0$, $P_1$, and $P_2$, which correlate with the membrane-level hypothesis of pore creation and expansion. These states differ in characteristic and relaxation times, contribution to increase in conductivity, and dependence with the applied electric field \cite{weaver1996theory,glaser1988reversible}. Although the model was built on the basis of the cell electroporation model of Leguèbe \textit{et al.}~\cite{Leguebe_2014} we proposed a different approach. Our model has an extra pore-detailing state (three states instead of two). In our definition, $P_0$ should explain the instantaneous increase in conductivity associated with rapid opening of hydrophobic pores (so-called prepores), $P_1$ should explain the initial formation of hydrophilic pores (initial pores), and $P_2$ the final expansion of the pores (final pores). Because of that, the relation between each state is also slightly different, $P_0$ and $P_1$ depend on the electric field, and $P_2$ depends directly on $P_1$.

We must note that the dynamics of electroporation is not yet fully understood~\cite{Scuderi_2022}, and there are several hypotheses on how the phenomenon occurs \cite{krassowska2007modeling,barnett1991electroporation,saulis1993celll,davalos2002feasibility, Mi_2022}. On the tissue scale, we can only assess a combined effect, which limits our conclusions about the physical changes on the membrane. Although we assume that each state is explained mainly by the aforementioned reasons, one would expect pores to form during $P_2$ and pores to expand during $P_1$, for example. Further work should be done to validate the correlations between tissue and membrane effects for each state.

The differences between the curves in Fig.~\ref{fig:beta} show that the saturation of $\beta_1$ is reached at electric field magnitudes of about half the saturation threshold of $\beta_0$. We believe that $\beta_1$ is saturated at lower electric field magnitudes due to the spatial distribution in the cell membrane for prepore and pore formation. Prepores are usually less stable and require less energy to form. Pores, on the other hand, will require more energy to form. For this reason, pores are likely to occur in the poles of the cell, where the electric field perpendicularly strikes its structure \cite{krassowska2007modeling, Milestone2022}. In low electric field stimulation, prepores and pores would likely form on the cell poles. Under these conditions, the rate of pore formation would be higher and a small amount of prepores would almost saturate the number of initial pores. Increasing the electric field magnitude would increase the number of prepores along the cell structure, but those would not form an initial pore. The difference in energy threshold would hamper the formation of pores throughout the cell structure. This could explain why increasing the magnitude of the electric field after a certain threshold keeps increasing the conductivity at the beginning of the pulse but does not change the increase amount over the pulse. As mentioned, the mechanics of single cell electroporation is not yet fully understood, and conclusions about its causes on the tissue scale are speculative.

The final pores cause larger flaws in the membrane than pre-pores or initial pores. Since the membrane is an insulating material, the flaw size leads the pre-pores to be less conductive than the initial and final pores. In fact, studies on cell suspensions under electroporation indicate an increase in conductivity over time during PEF stimulation, consistent with the higher conductivity of the final pores~\cite{suzuki2010theoretical}. There is a lack of information on conductivity during initial pore formation. Pre-pores are expected to form spontaneously even if the cell is at resting potential~\cite{weaver1996theory}. However, our proposed model does not consider the absolute values, but the concentration of the individual states. This consideration explains why the conductivity of the pre-pores state ($\sigma_{P_0}$) is greater than that of the initial pores ($\sigma_{P_1}$) and the final pores ($\sigma_{P_2}$). In absolute terms, we expect the number of pre-pores to be greater than the number of pores and final pores ($P_0 \gg P_1 \gg P_2$), but the concentration analysis normalises these values. Therefore, the average increase in conductivity is reflected in a higher value for the conductivity coefficient of the pre-pores than for the initial and final pores ($\sigma_{P_0} \gg \sigma_{P_1} \gg \sigma_{P_2}$).

The higher number of prepores formed in the first moments of the pulse accelerates the formation of the initial pores. This effect could be linked to theories that pores may be formed and expanded by coalescence of prepores \cite{sugar1987model,freeman1994theory}. Therefore, a higher concentration of $P_0$ would lead to a faster increase in $P_1$, as introduced in Eq.~\ref{eqn:tau-1}. $P_1$ and $P_2$ are also related. Here, we consider that the expansion of pores is proportional to the occurrence of initial pores. The difference is that the pores would take longer to expand and then longer to close. The timing of opening and closing of the pores should also be taken into account. There are differences in the mechanisms of opening and closing pores~\cite{yao2017analysis}. The mechanism of closing the pores is slower than that of opening them. Eqs.~\ref{eqn:tau-1} and \ref{eqn:tau-2} adjust the characteristic or relaxation time of states $P_1$ and $P_2$ depending on whether it is a growing or a decaying curve.

Fig.~\ref{fig:electric-currents} shows that our proposed model can describe the dynamics of the electric current during electroporation for tested input voltages. The solid foundation of the biological dispersion can be observed when \qty{10}{\kilo\volt\per\meter} (\qty{50}{\volt}) is applied (Fig.~\ref{fig:electric-currents}a). At \qty{10}{\kilo\volt\per\meter}, the influences of electroporation are small, so the electric current is explained mainly by biological dispersion (see that there is no increase in apparent conductivity in Fig.~\ref{fig:thermal-and-conductivity}b). Electroporation phenomena start to visually occur above \qty{20}{\kilo\volt\per\meter} (Fig.~\ref{fig:electric-currents}b), when the electric current deviates from the natural waveform of the biological dispersion. This threshold for the occurrence of electroporation is assessed in the curves of Fig.~\ref{fig:beta}, especially in the first dynamic dependence $\beta_0$. The $\beta_0$ shape is similar to the static model of \textit{Solanum tuberosum} proposed by Ivorra \textit{et al.}~\cite{Ivorra_2009}. The authors have developed a static model applying a single pulse \qty{400}{\micro\second} long and evaluating instantaneous conductivity at \qty{100}{\micro\second}. The similarity between $\beta_0$ and the static model is consistent, as $P_0$ has the greatest influence on the increase in tissue conductivity for the most magnitudes.

The results of the thermal analysis in Fig.~\ref{fig:thermal-and-conductivity}a show that the ESOPE protocol has a minimal temperature effect even at the higher electric field. The maximum increase was only \qty{0.88}{\celsius}, which is reflected in a change in conductivity of approximately \qty{1.5}{\percent}. Although we did not perform an experimental analysis of the thermal rise, our thermal simulation results are similar to those of other studies for the first eight pulses~\cite{Neal_2012, Suarez_2014, Zhao_2020}.

Fig.~\ref{fig:states} explains the electroporation dynamics for \qty{20}{\kilo\volt\per\meter} and \qty{100}{\kilo\volt\per\meter}. We can see that $P_1$ and $P_2$ concentrations increase from \qty{20}{\kilo\volt\per\meter}. As we defined that $P_1$ and $P_2$ are the states of hidrophylic pore formation, these states are expected to have the main influence on the increase in tissue permeability for macromolecules. Thus, cell permeability has already increased significantly at initial thresholds, where the effects of electroporation are not completely saturated. This confirms the statements in preclinical electrochemotherapy simulations that drug delivery is achieved with similar reliability after a certain threshold value of the electric field (the so-called reversible electroporation threshold)~\cite{Orlowski_1988, Rols_1990, Mahmood_2011, Boc_2018, Cindric_2022}. In further research, our aim is to better understand the effects of dynamic states on tissue permeability.

The increase in apparent conductivity with the influence of thermal and pore formation dynamics is shown in Fig.~\ref{fig:thermal-and-conductivity}b. We can observe that $P_1$ and $P_2$ are more influential up to \qty{30}{\kilo\volt\per\meter}, while $P_0$ significantly influence the increase in conductivity for thresholds higher than \qty{40}{\kilo\volt\per\meter}. As previously mentioned, this phenomenon may be related to the formation of pre-pores throughout the cellular structure that do not have sufficient energy to form a final pore under higher-intensity stimuli. Thus, the states $P_1$ and $P_2$ determine the increase in conductivity for thresholds below \qty{30}{\kilo\volt\per\meter}, while $P_0$ has a greater influence on stimuli above \qty{40}{\kilo\volt\per\meter}.

We found characteristic times for the first two dynamics similar to Voyer \textit{et al.}~\cite{Voyer_2018}. The authors analysed the electric current during the first pulse and used a similar set of equations to describe the first two dynamics. As only one pulse was analysed, they do not implement the relaxation between pulses or the third dynamic, which limits their model to a single-step analysis. In terms of increased conductivity, our results follow the same magnitude as those found by the static model of \textit{Solanum tubersum} by Ivorra \textit{et al.}~\cite{Ivorra_2009} and the dynamic model of Weinert \textit{et al.}~\cite{Weinert_2021} evaluated in rabbit tissues. Weinert \textit{et al.} compressed all electroporation dynamics into a single differential equation, which led to distinct dynamics of the increase in conductivity. We suspect that this compress resulted in parameter adjustments for voltage variations in their model (see Table 2 in~\cite{Weinert_2021}). A common factor among the three dynamic models of tissue electroporation proposed to this date is the adjustment of parameters based on input variations~\cite{Langus_2016, Voyer_2018, Ramos_2018, Weinert_2019, Weinert_2021}. In this sense, our model can accurately describe a wide range of input voltages with a single set of parameters.

The largest differences between the experimental average and the simulation results occur at \qty{20}{\kilo\volt\per\meter} (Fig.~\ref{fig:electric-currents}b) and \qty{50}{\kilo\volt\per\meter} (Fig.~\ref{fig:electric-currents}e). The difference at \qty{20}{\kilo\volt\per\meter} arises because the simulated increase in conductivity is slightly faster than that observed experimentally. We could implement a new function for the characteristic time of $P_1$ to account for this difference. However, this would introduce new parameters to fit only the first two pulses of an input electric field. For simplicity, we prefer not to make this consideration. The difference at \qty{50}{\kilo\volt\per\meter} is due to an overestimation of $\beta_0$ for this particular electric field. At \qty{50}{\kilo\volt\per\meter}, we are in the transition region of $\beta_0$, where small deviations in parameter definition would lead to large differences in electric current. It would be possible to better fit the result for \qty{50}{\kilo\volt\per\meter} while increasing the deviation for the others. Since the result for \qty{50}{\kilo\volt\per\meter} is at the upper edge of the standard deviation and all other input electric fields are close to the experimental average, we considered this set of parameters as the best choice.

\section{Conclusion}

We have proposed a novel dynamic model that describes tissue dielectric properties during PEF while accounting for electroporation, tissue dispersion, and temperature. The model divides the electroporation phenomenon into three dynamic states: prepore, initial pore, and final pore formation. The states are associated with microscopic effects on the membrane. The model can accurately describe the electric current during PEF. We believe that our proposed model can improve the study of PEF for electroporation-based applications.

\section*{Acknowledgements}
    This study was financed in part by \textit{Coordenação de Aperfeiçoamento de Pessoal de Nível Superior -- Brasil} (CAPES) and \textit{Conselho Nacional de Desenvolvimento Científico e Tecnológico -- Brasil} (CNPq).

\section*{Author Declarations}

    \subsection*{Conflict of Interest}
        The authors have no conflicts of interest to disclose.

    \subsection*{Author Contributions}
        \textbf{R.G.}: Conceptualisation, Methodology, Formal analysis, Investigation, Data Curation, Writing – Original Draft, Writing – Review \& Editing, Visualisation. \textbf{D.L.L.S.A.}: Methodology, Investigation, Data Curation, Writing – Review \& Editing. \textbf{J.R.S.}: Formal analysis, Investigation, Writing – Review \& Editing. \textbf{G.B.P.}: Resources, Supervision, Writing – Review \& Editing. \textbf{D.O.H.S.}: Resources, Supervision, Project administration, Writing – Review \& Editing.

\section*{Data Availability}
    The data that support the findings of this study are available from the corresponding author upon reasonable request.

\section*{References}
\bibliography{references}% Produces the bibliography via BibTeX.

\end{document}